**Spin Electronics**___________________________________________________________________________

# Pulse Shaping to Mitigate the Impact of Device Imperfections in Field-Free Switching Using Combined Spin-Orbit and Spin-Transfer Torques

Kuldeep Ray[1,2], Jérémie Vigier[1,2], Sylvain Martin[2], Chloé Bouard[2], Nicolas Lefoulon[2], Marc Drouard[2], and Gilles Gaudin[1]

[1] Université Grenoble Alpes, CEA, CNRS, Grenoble-INP, SPINTEC, 38000 Grenoble, France
[2] Antaios, 38240 Meylan, France

*Abstract*—Combining spin-orbit (SOT) and spin-transfer torques (STT) provides a practical approach for field-free switching in spin-orbit torque magnetic random-access memory (SOT-MRAM), a prerequisite for industrial deployment, but can compromise reliability through phenomena such as backhopping, especially in top-pinned stacks commonly used for SOT-MRAM. We investigate the write error rate (WER) of combined SOT + STT switching in top-pinned devices that are not optimized for STT switching. Experiments reveal clear indications of STT-induced backhopping and a pronounced field-free SOT switching asymmetry between AP-to-P and P-to-AP transitions. Our macrospin model, using two coupled Landau Lifshitz Gilbert equations for the free and the reference layers, qualitatively reproduces this asymmetry and reveals an intermediate loss-of-determinism regime in addition to the well-known backhopping region. Based on these simulations, we propose mitigation strategies and experimentally demonstrate that STT pulse shaping reduces WER and improves switching robustness in the presence of device imperfections.

## I. INTRODUCTION

The scaling of complementary-metal-oxide-semiconductor (CMOS) technology in recent years has significantly increased power consumption in computing units. Integration of non-volatile memory technologies close to the central processing unit (CPU) could potentially reduce power consumption. Spin-orbit torque magnetic random-access memory (SOT-MRAM) is one of the leading candidates for cache-level CMOS replacement, particularly due to the separate read and write paths allowing enhanced reliability and fast operation speeds [Prenat 2016, Nguyen 2024, Krizakova 2022]. However, several challenges must be tackled before SOT-MRAM integration into functional memory arrays.

A key challenge is the need for a symmetry-breaking mechanism to achieve deterministic switching [Nguyen 2024, Krizakova 2022]. Although initial demonstrations used externally applied magnetic fields, this approach is incompatible with embedded memory applications. To tackle this issue, numerous approaches have been proposed, including interlayer exchange coupling [Liu 2019], free layer with tilted anisotropy [You 2015], exchange bias [van den Brink 2016], combining SOT and spin-transfer torque (STT) effects [Cai 2021] and unconventional torques from novel materials [Kao 2022].

Most of these approaches, particularly those relying on novel materials, would require significant process modifications for large-scale integration. In contrast, field-free switching based on combined SOT and STT presents a more practical solution, as it requires fewer material and process modifications. Typically, STT switching needs an incubation time to drive large magnetization precessions eventually resulting in switching. When combined, SOT reduces this incubation time by providing an initial magnetization tilt at low voltages or driving it in-plane at high voltages, while STT determines the switching direction. Nevertheless, incorporating STT in SOT-MRAM can also introduce error mechanisms typical of STT-MRAM, including backhopping [Devolder 2020]. STT-MRAM typically uses a bottom-pinned magnetic tunnel junction (MTJ), i.e., the pinning layer is at the bottom followed by the reference layer, the oxide barrier, and the free layer on top. Conversely, SOT-MRAM most commonly employs top-pinned MTJs, but they tend to exhibit more structural imperfections compared to bottom-pinned MTJs [Swerts 2017], thereby increasing the likelihood of switching failures and write errors.

Previous works combining SOT and STT pulses [Wang 2015, Pathak 2020, Cai 2021] either used identical SOT and STT pulse widths or a short SOT with a longer synchronous STT pulse. The impact of device imperfections on field-free switching robustness has been largely ignored. In this study, we investigate field-free switching in SOT-MRAM using combined SOT and STT pulses via the write error rate (WER) measurements. We report experimental signatures consistent with device imperfections in our stacks. Using macrospin simulations incorporating two coupled Landau-Lifshitz-Gilbert (LLG) equations describing the dynamics of the free and reference layers, we qualitatively analyze the impact of these imperfections on magnetization switching using combined SOT and STT. Based on this analysis, we propose several methods to improve WER and widen the deterministic switching window. Additionally, we experimentally validate one such method based on STT pulse-shape engineering, demonstrating improved WER and robust field-free switching even in

___________________________________________________________________________________________________________________

the presence of the identified imperfections.

## II. METHODOLOGY

### A. Sample Description

The measured devices consist of a 65 nm diameter top-pinned MTJ pillar patterned on a 130 nm wide SOT track and fabricated on 300 mm wafers. The material stack (bottom to top) is shown in Fig. 1(a): β-W(4)/Co$_{20}$Fe$_{60}$B$_{20}$(0.9)/MgO(RA~20 Ωμm$^2$)/Co$_{20}$Fe$_{60}$B$_{20}$(1.0)/Ta(2)/[Co(0.25)/Pt(0.3)]$_3$/Co(0.4)/Ir(0.45)/Co(0.4)/[Pt(0.3)/Co(0.21)]$_8$/Pt(0.3)/Ta(1), where the thicknesses are expressed in nanometers. The 0.9 nm CoFeB free layer, with perpendicular magnetic anisotropy (PMA), exhibits a coercivity of 47.75 kA/m and an offset field of H$_{off}$ = –11.94 kA/m (Fig. 1(b)).

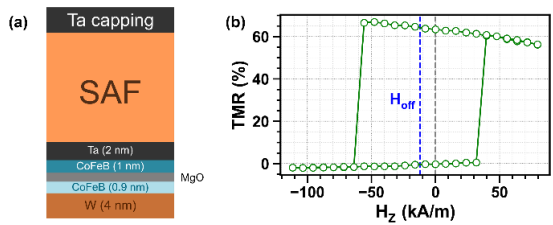

Fig. 1. (a) Schematic of the material stack of SOT-MRAM devices measured in this study. (b) Representative TMR hysteresis loop obtained by sweeping the out-of-plane field (H$_Z$), showing a TMR = 63% at H$_Z$=0 and an offset field H$_{off}$ = –11.94 kA/m.

### B. Experimental Setup

The measurement setup is shown in Fig. 2(a). An arbitrary waveform generator delivers independent nanosecond SOT (channel 1) and STT (channel 2) voltage pulses to the SOT track and the MTJ terminals, respectively, via the rf ports of two bias tees. For readout, a separate source applies microsecond voltage pulses via the bias tee dc port. The rf component of the transmitted voltage is monitored on an oscilloscope, while the dc component is sent to an external 3 kΩ load. The voltage drop measured across this load depends on the parallel or antiparallel alignment of the MTJ free and reference layers and is proportional to the tunnel magnetoresistance (TMR).

For switching with only SOT pulses, we apply a compensation voltage (0.65·V$_{SOT}$) at the MTJ terminal to ensure no STT current. For switching with only STT pulses, a similar compensation voltage (0.09·V$_{STT}$) is applied to the SOT terminal such that the current is equally divided between both arms of the SOT track.

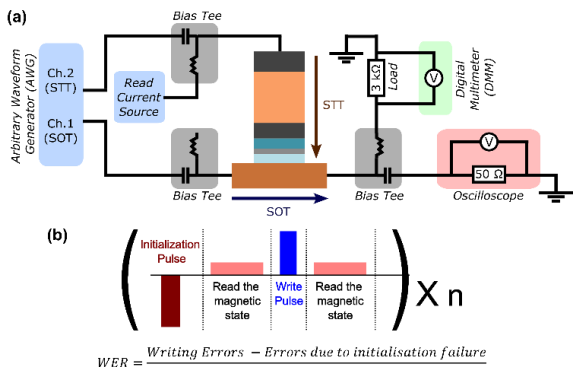

Fig. 2. (a) Schematic of the experimental setup. (b) WER definition and measurement scheme.

### C. Write Error Rate

The statistical measurement method of write error rate (WER) is extensively used in this work to characterize switching probability in SOT-MRAM devices. It is a key metric for memory devices, typically required to be below 10$^{-9}$ in memory arrays with error correction code (ECC) or below 10$^{-18}$ in the absence of ECC [Khvalkovskiy 2013].

In this method, a first pulse with fixed amplitude defines the initial state of the MTJ (either parallel (P) or antiparallel (AP)), and the magnetization state is read to verify that the initialization was successful. A pulse with a certain amplitude is then applied to write the opposite state, and the magnetic state is read again to verify that the opposite state has been written. This cycle is repeated several times, and the WER for a certain amplitude of write pulse is defined as the ratio of the number of writing errors after a successful initialization to the number of successful initializations, as shown in Fig. 2(b). Repeating this process for different pulse amplitudes yields WER curves plotted on a logarithmic scale as a function of pulse amplitude. Such representation emphasizes writing errors that are often hidden in linear switching probability plots. It is important to note that a WER of 0.5 indicates an equal probability of obtaining the P or AP state.

## III. RESULTS AND DISCUSSION

We first characterized SOT-only switching using 5 ns-long SOT pulses in the presence of an external in-plane magnetic field of H$_X$ = 63.66 kA/m. The measured WER curves, such as that shown in Fig. 3(a), show deterministic switching with a WER < 10$^{-5}$ for both AP-to-P and P-to-AP transitions without any SOT-induced backswitching. With an in-plane assist field, SOT switching is well-behaved and yields low WER for both polarities. By contrast, in the absence of an external field, an ideal device would exhibit overlapping AP-to-P and P-to-AP WER curves that approach WER = 0.5 at large |V$_{SOT}$|. Our measurements clearly deviate from this expectation. As seen in Fig. 3(b), the AP-to-P threshold is shifted by ~ 0.4 V relative to P-to-AP. Additionally, the P-to-AP WER curve demonstrates a peculiarity that is absent for AP-to-P. A similar asymmetry has been detected in all measured devices, despite device-to-device variability. This significant asymmetry suggests the presence of magnetic fields internal to the device, such as the stray fields from the MTJ stack or additional interlayer couplings (for example, exchange coupling or orange-peel coupling induced by MgO roughness) [Yang 2021].

We next characterized STT-only switching in SOT-MRAM devices in the absence of an external magnetic field using 5 ns and 10 ns pulses, as shown in Fig. 3(c). Since STT switching is deterministic on its own, lower WER values are obtained. However, the measured WER remains well above optimized STT-MRAM [Nakayama 2023]. A clear AP-to-P / P-to-AP asymmetry further supports the presence of internal magnetic fields. Additionally, the well-known phenomenon of STT backhopping, characterized by an abrupt jump in WER at high STT voltages, is seen for both transitions using a 10 ns-long STT pulse. Similar backhopping is detected in all measured devices, which is not surprising since the stacks are optimized for field-assisted SOT switching rather than for STT switching. One potential explanation for backhopping is an insufficiently stable reference layer caused by weak pinning. In top-pinned SOT-MRAM, the synthetic antiferromagnetic (SAF) layer is grown on top of a (100)-textured CoFeB/MgO/CoFeB stack [Swerts 2017], which is less favorable for achieving high-quality (111)-textured pinning layer compared with bottom-pinned MTJs, where the SAF layer is grown first on a (111)-textured seed layer, resulting in improved pinning and thermal stability [Liu 2017]. Such non-ideal growth can result in weak pinning and low stability of the reference layer [Tomczak 2016].

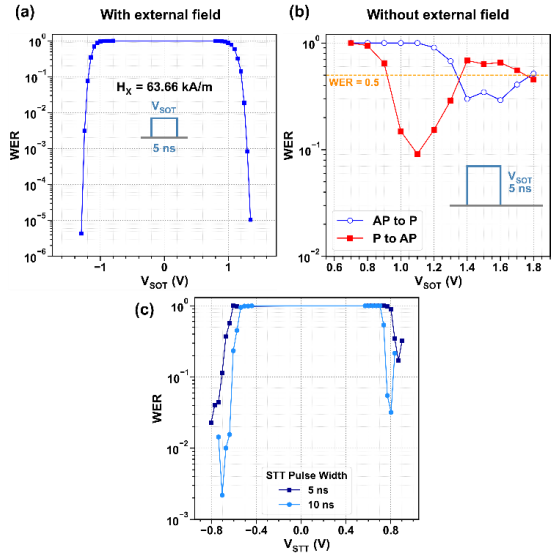

Fig. 3. WER as a function of SOT voltage ($V_{SOT}$): (a) in the presence of an external in-plane magnetic field, $H_X$ = 63.66 kA/m, and (b) without any external magnetic field. (c) WER as a function of STT voltage ($V_{STT}$) for pulse widths of 5 ns and 10 ns.

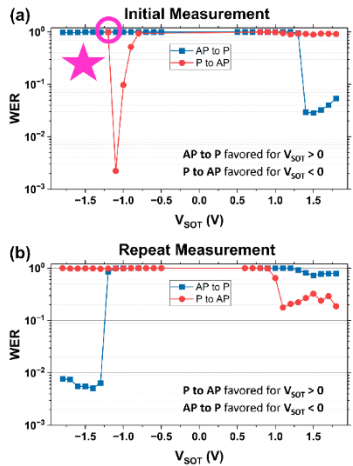

Fig. 4. (a) Field-free WER as a function of SOT voltage ($V_{SOT}$). (b) Repeat measurement on the same device under identical conditions.

Additional evidence for imperfections is provided by an irreversible change in device behavior following a potential backhopping event. We measured WER as a function of $V_{SOT}$ for both P-to-AP and AP-to-P transitions for positive and negative SOT voltages, as shown in Fig. 4. In these measurements, an STT initialization pulse with amplitude above the STT switching threshold is used to set the initial state. In the first measurement (Fig. 4. (a)), the internal magnetic fields favor the P state for $V_{SOT} > 0$ and the AP state for $V_{SOT} < 0$. However, we detected an abrupt jump in WER for $V_{SOT}$ = –1.2 V that is uncharacteristic of SOT switching in the absence of an external field. We attribute this to a disturbance of the reference layer (or another layer in the top stack) induced by the high-amplitude STT initialization pulse, consistent with prior reports on STT-driven instabilities [Devolder 2020]. We repeated the same measurement on the same device under identical conditions to test this hypothesis. A completely different behavior was indeed observed. In the repeated measurement, the internal fields now favor the AP state for $V_{SOT} > 0$ and the P state for $V_{SOT} < 0$, which is opposite to the initial measurement. This suggests a significant change in the reference layer or the MTJ top stack and reinforces the picture of low stability and weak pinning.

## IV. MACROSPIN SIMULATIONS

We performed macrospin simulations incorporating interlayer coupling and reference layer defects to study the impact of such device imperfections on field-free switching using combined SOT and STT pulses. For MTJ diameters above 20 nm - 30 nm, SOT-driven reversal is most accurately captured by micromagnetic simulations. However, they are computationally intensive and therefore impractical for generating statistically meaningful WER curves. In contrast, macrospin simulations provide a faster alternative and have been shown to effectively reproduce experimental WER curves [Ray 2025]. Here, we numerically solved two coupled LLG equations describing the free and reference layers, incorporating additional terms for STT, as well as damping-like and field-like SOT. The dynamics of the free layer magnetization ($\boldsymbol{m_{free}}$) are given by:

$$\frac{d\boldsymbol{m_{free}}}{dt} = -\gamma\mu_0(\boldsymbol{m_{free}} \times \boldsymbol{H_{eff}}) + \alpha_{free}\left(\boldsymbol{m_{free}} \times \frac{d\boldsymbol{m_{free}}}{dt}\right) \quad (1)$$
$$-\gamma\mu_0(\boldsymbol{m_{free}} \times \boldsymbol{H_{FL}}) - \gamma\mu_0(\boldsymbol{m_{free}} \times \boldsymbol{H_{DL}})$$
$$-\gamma\mu_0(\boldsymbol{m_{free}} \times \boldsymbol{H_{STT}})$$

where $\boldsymbol{H_{eff}}$ is the effective field including the anisotropy field and exchange coupling field, $\boldsymbol{H_{DL}} = \frac{j_{SOT}\theta_{SH}\hbar}{2\mu_0 e t_{free} M_{S,free}}(\boldsymbol{m_{free}} \times \boldsymbol{u_Y})$ is the effective damping-like term, $\boldsymbol{H_{FL}} = \frac{\beta j_{SOT}\theta_{SH}\hbar}{2\mu_0 e t_{free} M_{S,free}} \boldsymbol{u_Y}$ is the effective field-like term, and $\boldsymbol{H_{STT}} = \frac{j_{STT}\eta\hbar}{2\mu_0 e t_{free} M_{S,free}}(\boldsymbol{m_{free}} \times \boldsymbol{m_{ref}})$ is the STT effective field term. The dynamics of the reference layer magnetization ($\boldsymbol{m_{ref}}$) are given by a similar equation excluding the SOT terms and $\boldsymbol{H_{STT}} = \frac{j_{STT}\eta\hbar}{2\mu_0 e t_{ref} M_{S,ref}}(\boldsymbol{m_{ref}} \times \boldsymbol{m_{free}})$. The simulation parameters are given in Table 1. The free and reference layer (RL) thicknesses were taken from the device stack. The free layer anisotropy field was measured. The free layer saturation magnetization, damping parameter, and the reference layer parameters were taken from the literature [Pathak 2020]. We set the spin Hall angle $\theta_{SH}$ = 0.25 and the field-like to damping-like ratio, $\beta = -1$, within reported ranges for W-based systems [Krizakova 2022]. To reflect the unoptimized STT response of our devices, the STT efficiency ($\eta = -0.21$) was kept low compared to the value of –0.6 used in previous studies [Pathak 2020, Wang 2015]. $\eta$, $H_{k,ref}$ and $\alpha_{ref}$ were adjusted to observe STT backhopping in our simulations. To generate WER curves, we included a random thermal field $\boldsymbol{H_{th}}$ in $\boldsymbol{H_{eff}}$, $\boldsymbol{H_{th}} = \sqrt{\frac{2\alpha k_B T}{\gamma \mu_0^2 V M_{S,free\,or\,ref} \Delta t}} \boldsymbol{\zeta_{th}}$ where $\boldsymbol{\zeta_{th}}$ is a random Gaussian unit vector. For each write condition, we simulated 100 writing attempts and computed WER.

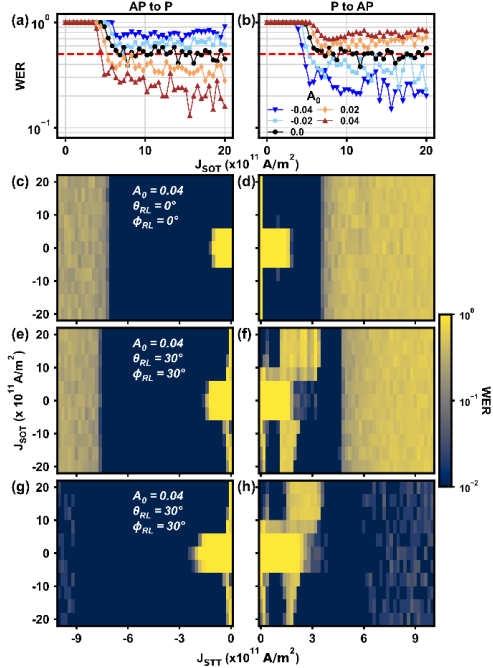

Fig. 5. Simulated WER versus SOT current density using different coupling parameters for (a) AP to P and (b) P to AP transitions. Simulated WER colormaps as a function of SOT and STT current densities for AP to P and P to AP transitions under different conditions: (c-d) perpendicular RL with $A_0>0$, (e-f) tilted RL with $A_0>0$, and (g-h) tilted RL with $A_0>0$ using a modified pulse (1 ns plateau and 9 ns decay).

Table 1. Simulation parameters and their respective values.

| | Parameter | Value |
|---|---|---|
| Free layer | Anisotropy field ($H_{k,free}$) | 238.73 kA/m |
| | Saturation magnetization ($M_{S,free}$) | $10^6$ A/m |
| | Magnetic damping constant ($\alpha_{free}$) | 0.03 |
| | Thickness ($t_{free}$) | 0.9 nm |
| Reference Layer | Anisotropy field ($H_{k,ref}$) | 636.62 kA/m |
| | Saturation magnetization ($M_{S,ref}$) | $10^6$ A/m |
| | Magnetic damping constant ($\alpha_{ref}$) | 0.01 |
| | Thickness ($t_{ref}$) | 1.1 nm |
| Other Parameters | Spin Hall angle ($\theta_{SH}$) | 0.25 |
| | Field-like to damping-like ratio ($\beta$) | -1 |
| | STT efficiency ($\eta$) | -0.21 |
| | MTJ diameter ($d$) | 30 nm |

To model coupling between free and reference layers, we added an exchange coupling term, $H_{ref \to free \, or \, free \to ref} = \frac{A_0 M_{S,ref} M_{S,free}}{M_{S,free \, or \, ref}} m_{ref \, or \, free}$, to $H_{eff}$. Fig. 5(a-b) shows WER curves for varying coupling coefficient ($A_0$). In the absence of both an external field and coupling, SOT switching is random, causing the WER to saturate at 0.5. When coupling is introduced, either the P-to-AP or the AP-to-P transition becomes favored, depending on $A_0$, and the WER saturates at different values. These results are qualitatively similar to experimentally measured field-free SOT WER, supporting the conclusion that interlayer coupling contributes to the observed behavior. Applying an in-plane symmetry-breaking field mitigates the impact of this coupling, restoring low and nearly symmetric WER.

Fig. 5(c-d) shows WER colormaps using 5 ns SOT and 10 ns STT pulses for an arbitrary coupling coefficient, $A_0 = 0.04$. Three distinct regions can be identified: (i) no-switching region (yellow) for low $J_{SOT}$ and $J_{STT}$; (ii) deterministic switching region (deep blue); and (iii) STT backhopping region at high $J_{STT}$. Since $A_0 = 0.04$ favors AP-to-P transitions, the deterministic switching region is wide in Fig. 5(c) and narrower in Fig. 5(d), whereas in the absence of coupling, both colormaps are identical. Finally, to emulate weak pinning and partial loss of PMA potentially caused by texture mismatch between the reference and SAF layers, we introduced an arbitrary tilt of the RL magnetization ($\theta_{RL} = 30°$, $\varphi_{RL} = 30°$). This RL tilt defect produces an additional intermediate loss-of-determinism regime (yellow bands embedded within the deterministic regions for $J_{SOT} > 0$ in P-to-AP switching, as seen in Fig. 5(f)). The position and width of these bands depend on the tilt orientation, and they vanish when the coupling is removed. These results suggest that this intermediate region arises from a combination of the RL tilt and coupling between the free and reference layers. Furthermore, a partial switching of RL could result in a change in $\theta_{RL}$ and $\varphi_{RL}$, altering the device behavior, as shown experimentally in Fig. 4. Improving the stack quality to minimize interlayer couplings and enhance PMA could profoundly improve switching performance.

## V. PULSE SHAPING TO REDUCE BACKHOPPING

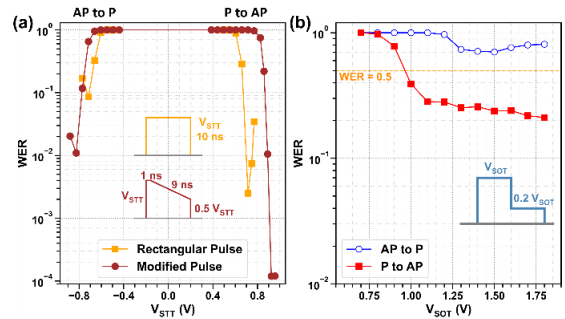

Fig. 6. (a) WER versus $V_{STT}$ for standard rectangular and modified pulse shapes. (b) WER versus $V_{SOT}$ for a modified two-step SOT pulse.

Despite the presence of imperfections, the simulations in Fig. 5(g-h) show suppression of the STT backhopping region using a modified STT pulse with a 1 ns plateau followed by a 9 ns linear decay. We experimentally validated this solution. As seen in Fig. 6(a), the WER for standard rectangular pulses (10 ns at $V_{STT}$) is limited to 0.003 for P-to-AP and to 0.09 for AP-to-P transitions, whereas using the modified pulses (1 ns at $V_{STT}$ followed by a 9 ns decay to $0.5 \cdot V_{STT}$) lowers the WER by at least one decade for both transitions. This improvement is systematically observed across all measured devices. We attribute this enhanced reliability to a more controlled energy dissipation during the shaped pulse that prevents excess energy accumulation in the RL and effectively mitigates backhopping. On the other hand, a rectangular pulse maintains a high $V_{STT}$ throughout the full 10 ns duration, injecting excess energy into the RL, promoting its destabilization and ultimately triggering backhopping. Although the shaped pulse requires a higher nominal switching voltage, the switching energies for both pulses are similar for P to AP and even lower for the modified pulse for AP to P transition.

We also observed that a two-step SOT pulse (5 ns at $V_{SOT}$ followed by 5 ns at $0.2 \cdot V_{SOT}$) suppresses the peculiar oscillatory behavior shown in Fig. 3(b) (Fig. 6(b)). Combining this modified SOT pulse with the modified STT pulse facilitates field-free switching without any backhopping, as shown in Fig. 7. Increasing $V_{STT}$ increases the slope of the WER curve, and even subcritical STT voltages ($V_{STT}$ = 0.66 V or 0.77 V) reduce the WER by more than an order of magnitude. Further increasing $V_{STT}$ to 0.88 V, where the STT begins to switch on its own, significantly improves WER performance, as shown in Fig. 7(a).

Finally, we fix $V_{STT}$ = 0.77 V and vary the temporal overlap between the SOT and STT pulses. As shown in Fig. 7(b), increasing the overlap improves WER: an overlap of 1 ns yields approximately one decade lower WER compared to zero overlap. Interestingly, introducing a separation between the SOT and STT pulses degrades WER such that the curve for $V_{STT}$ = 0.77 V and -2 ns delay in Fig. 7(b) is identical to the curve without STT in Fig. 7(a). These results emphasize the importance of both the STT amplitude and the SOT-STT overlap in achieving low-error, field-free switching. Further improvements will ultimately require material and process optimization to reduce device imperfections, as well as a more systematic exploration of the pulse-parameters space for combined SOT + STT operation.

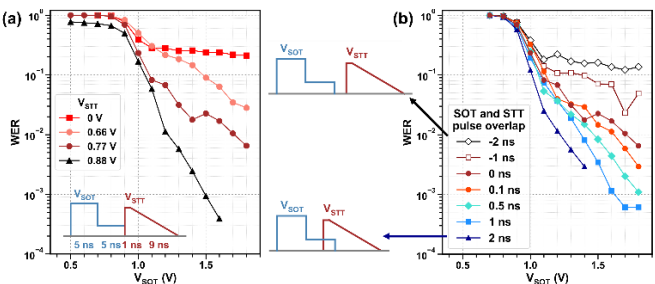

Fig. 7. WER versus $V_{SOT}$ using modified SOT and STT pulses (shown in inset) (a) for different $V_{STT}$, and (b) for different overlaps between SOT and STT pulses with $V_{STT}$ = 0.77 V. A negative overlap corresponds to the separation between the end of the SOT and the beginning of the STT pulses.

## VI. CONCLUSION

WER measurements provide substantial evidence of device imperfections in our top-pinned SOT-MRAM. Through macrospin simulations, we clarify how these imperfections affect combined SOT+STT switching, leading not only to the well-known STT backhopping regime but also to an additional intermediate loss-of-determinism region. Guided by the model, we numerically and experimentally demonstrate that shaping both the STT and SOT pulses significantly reduces WER. Using the modified pulse scheme, we achieve field-free switching with improved reliability, offering a viable pathway toward robust magnetization switching via combined SOT and STT mechanisms.

## ACKNOWLEDGMENT

We thank the staff of Antaios for their support throughout this project. This project has received funding from the European Union's Horizon 2020 research and innovation programme under the Marie Skłodowska-Curie Grant Agreement No. 955671, and was supported by the Région Auvergne-Rhône-Alpes Pack Ambition Recherche Program (Grant No. 19-009938-01-MAPS).